\documentclass[showpacs,preprintnumbers,twocolumn]{revtex4}
\usepackage{amsmath,amssymb,pstricks,wasysym,stmaryrd,enumerate,epsfig}
\begin{document}
\title{CP-violating theta parameter in the domain model of the QCD vacuum}
\author{Alex C. Kalloniatis\footnote{akalloni@physics.adelaide.edu.au} }

\address{
Special Research Centre for the Subatomic Structure of Matter,
University of Adelaide,
South Australia 5005, Australia}

\author{Sergei N. Nedelko \footnote{nedelko@thsun1.jinr.ru}}
\address{ Bogoliubov Laboratory of Theoretical Physics, JINR,
141980 Dubna, Russia}

\date{\today}
\preprint{ADP-04-27/T609}
\begin{abstract}
A non-zero CP-violating $\theta$ parameter is treated in the domain model
which assumes a cluster-like vacuum structure whose
units are characterised in particular by a topological charge which is 
not necessarily an integer number.
In the present paper we restrict consideration to rational values of the 
charge.
The model has previously been shown to manifest
confinement, spontaneous chiral symmetry breaking and
the absence of an axial $U(1)$ Goldstone boson.
We find that the specific structure of the minima of the free energy 
density of the domain ensemble forces
a $2\pi$-periodicity of observables in $\theta$  
for any number of light quarks, that vacuum doubling occurs
at $\theta=\pi$ for any $N_f>1$ and any value of topological charge $q$.
These features are in agreement with expectations
based on anomalous Ward identities and large $N_c$
effective theories.
We find also  
additional values of $\theta$ depending on $q$ for which vacuum
doubling occurs. 
\end{abstract}
\maketitle

\section{Introduction}
Explicit CP violation can be introduced in
quantum chromodynamics by the inclusion 
 of the so-called  theta term in the action.
In Euclidean space this amounts to
\begin{eqnarray}
\label{thterm}
S_\theta=iq\theta, \ \ 
q= {{g^2}\over{32 \pi^2}} \int d^4x F_{\mu \nu} {\tilde F}^{\mu \nu} .
\end{eqnarray}
A remarkable feature of such
a term in the presence of spontaneously broken chiral
symmetry is the Dashen phenomenon \cite{Das71}:   
this explicit breaking of CP can become  spontaneous
due to vacuum doubling  at the point $\theta=\pi$. 
This value falls outside the physically relevant range
$0\leq\theta\leq 10^{-9}$ to which $\theta$ is constrained
by the neutron dipole moment \cite{thetameasure} and current
algebra \cite{CdVVW79}.  Nonetheless, the Dashen phenomenon
is a constraint
on self-consistent models of the non-perturbative QCD vacuum. 
The statement of the problem related to the Dashen phenomenon as well  
as the question about periodic dependence of observables in $\theta$ 
acquire full sense only if the values of $q$ are not 
restricted to integers.
In this paper we use the domain model \cite{NK2001} of the QCD vacuum to
show how specific properties of quark field configurations summed up in 
the partition function
can lead to the periodic dependence on $\theta$ and to the Dashen phenomenon 
for any rational values of $q$. 

The approaches which originally demonstrated the Dashen
phenomenon in the context of QCD are those of
anomalous Ward identities \cite{CrewNATO} and effective
chiral Lagrangians \cite{largeNc} in the limit of large number of colours
$N_c$,
which we shall describe below in more detail. Subsequent discussions
of the Dashen phenomenon include 
\cite{Palmer:1980,Crewther:1982,LeuSmil92,ALS02,Creutz03}.
This phenomenon for $\theta\neq 0$ is intimately related to both 
the mechanism of spontaneous chiral symmetry breaking
and the non-appearance of an axial $U(1)$ pseudoscalar
Goldstone boson as shown by the original works revealing
it in QCD. In \cite{CrewNATO} the key tool is
the generalisation to the axial $U(1)$ channel of
Ward identities in which the true vacuum of
the theory is  unknown, but for which divergences of
current-current  expectation values in this vacuum
can be related to hadron spectroscopy via vacuum symmetry properties.  
For example, the light pion mass as input into the 
flavour $SU(2)_L \times SU(2)_R$ Ward identity
(which gives the Gell-Mann--Oakes--Renner relation) 
leads to a non-zero chiral condensate
as the output. In the axial $U(1)$ channel  
(where the anomaly figures) and $\theta=0$ the phenomenological input
is now the absence of a light meson,
with the output that topological charge $q$ must be fractional.
With $\theta \neq 0$ the corresponding output starting with the
same phenomenological input is $2\pi$-periodicity in $\theta$
and the property of vacuum doubling and spontaneous CP breaking
at $\theta=\pi$. In the large $N_c$ approach the same
data is turned around. An effective Lagrangian for mesons is written
down for $N_c\rightarrow \infty$ which assumes 
chiral symmetry breaking and the anomaly.
The output for $\theta=0$ includes the famous relation
between the topological susceptibility and various
meson masses. For $\theta \neq 0$ again the Dashen phenomenon
emerges.

However these approaches do not unveil
the actual mechanism of non-perturbative vacuum properties
such as confinement and spontaneous chiral symmetry breaking.
The domain model \cite{NK2001} is a bottom up approach to these features:
a particular vacuum structure is introduced into the formalism explicitly, 
here based on a statistical ensemble of domain-like gluon fields, 
and out of this both vacuum and mesonic properties are derived. 
In previous works both confinement \cite{NK2001}
and chiral symmetry realisation \cite{NK2002, NK2004} have been 
studied for this vacuum {\it ansatz}. 
The anomaly contribution to the
free energy suppresses continuous axial $U(1)$ degeneracy in
the ground state, leaving only
a discrete residual axial symmetry. This discrete symmetry 
and flavour $SU(N_f)_L\times SU(N_f)_R$ chiral symmetry
in turn are spontaneously broken
with a quark condensate arising due to the asymmetry
of the spectrum of Dirac operator in the domain background.  
An estimate of pseudoscalar and vector meson masses has been computed 
showing  the typical pattern of spontaneous breakdown of  
$SU(N_f)_L\times SU(N_f)_R$ chiral symmetry
 with the correct splitting between 
$\pi,\eta$ and $\eta^\prime$ mesons \cite{NK2004}.
In this paper we test the model against the above predictions of
spontaneous CP breaking and generally explore its features for non-zero
$\theta$ parameter. The Dashen phenomenon is studied by 
directly computing the $\theta$-dependendence of the free energy density 
and chiral condensate within the domain model.

The results of this paper are that indeed the
vacuum doubling at $\theta=\pi$ occurs in the domain model for
any
rational values of the topological charge $q$ and any  
number of flavours.
Thus the domain model provides an explicit example of  
simultaneous realisation of
confinement and chiral symmetry breaking where the Dashen phenomenon
is manifest.  Within the model, the topological charge is not restricted to any 
{\it ad hoc} value, and in the general case there are critical values
in addition to $\theta=\pi$, which disappear
when $q$ is set to $1/N_f$ (which is the value of $q$ emerging in 
\cite{CrewNATO} for the case of $N_f$ degenerate quark flavours).
In particular, in \cite{NK2001} the domain model parameters were fixed
from the string tension giving $q=0.15$.
For this topological charge, the critical values
$\theta=-\pi/3,\pi/3,\pi({\rm mod} 2\pi)$
appear for which the vacuum doubling occur.  
As with \cite{CrewNATO}, vacuum degeneracy for discrete $\theta$ and  
$2\pi$-periodicity of observables such as the condensate
 arises as output rather than being assumed. 
 Unlike \cite{CrewNATO,largeNc}, these results emerge without
the absence of a massless $U(1)$ boson being input.

We first briefly review the 
domain model for $\theta=0$ and thereafter examine the
realisation of chiral symmetry for $\theta \neq 0$. In the appendices
we review and compare our results with the approaches of \cite{CrewNATO}
and \cite{largeNc}.
 
\section{Review of chiral symmetry in the domain model}
For motivation and a detailed description of the domain model we refer the 
reader to \cite{NK2001}.
The essential definition of the model
is given in terms of the following partition function for
$N\to\infty$ domains of radius $R$
\begin{eqnarray}
{\cal Z} & = & {\cal N}\lim_{V,N\to\infty}\int_{-\infty}^{\infty} d\alpha
\prod\limits_{i=1}^N
\int\limits_{\Sigma}d\sigma_i
\int_{{\cal F}_\psi^i}{\cal D}\psi^{(i)} {\cal D}\bar \psi^{(i)}
\nonumber \\
&&\times \int_{{\cal F}^i_Q} {\cal D}Q^i 
\delta[D(\breve{\cal B}^{(i)})Q^{(i)}]
\Delta_{\rm FP}[\breve{\cal B}^{(i)},Q^{(i)}]
\nonumber \\
&&\times e^{
- S_{V_i}^{\rm QCD}
\left[Q^{(i)}+{\cal B}^{(i)}
,\psi^{(i)},\bar\psi^{(i)}
\right]}
\label{partf}
\end{eqnarray}
where the functional spaces of integration
${\cal F}^i_Q$ 
and ${\cal F}^i_\psi$  are specified by the boundary conditions
$(x-z_i)^2=R^2$
\begin{eqnarray}
\label{bcs}
&&\breve n_i Q^{(i)}(x)=0, 
\nonumber\\
&&i\!\not\!\eta_i(x) e^{i\alpha\gamma_5}\psi^{(i)}(x)=\psi^{(i)}(x),
\label{quarkbc} \\
&&\bar \psi^{(i)} e^{i\alpha\gamma_5} i\!\not\!\eta_i(x)=-\bar\psi^{(i)}(x).
\nonumber
%\label{adjquarkbc} 
\end{eqnarray}
Here $\breve n_i= n_i^a t^a$ with the 
generators $t^a$  of $SU_{\bf c}(3)$ in the adjoint representation
and $\alpha$ is a random chiral angle
associated with the chiral symmetry violating boundary condition
Eq.(\ref{quarkbc}) in the presence of one species of quark.
The generalisation to $N_f$ flavours is given in the next section. 
The thermodynamic limit assumes $V,N\to\infty$ but 
with the density $v^{-1}=N/V$ taken fixed and finite. The
partition function is formulated in a background field gauge
with respect to the domain mean field, which is approximated 
inside and on the boundaries of the domains by
a covariantly constant (anti-)self-dual gluon field with the 
field-strength tensor of the form
\begin{eqnarray*}
F^{a}_{\mu\nu}(x)
=
\sum_{j=1}^N n^{(j)a}B^{(j)}_{\mu\nu}\vartheta(1-(x-z_j)^2/R^2), 
\end{eqnarray*}
with $B^{(j)}_{\mu\nu}B^{(j)}_{\mu\rho}=B^2\delta_{\nu\rho}$.
Here  $z_j^{\mu}$ are the positions of the centres of domains in 
Euclidean space.

In order to reflect continuity of colorless quark currents at the 
positions of pure gauge singularities 
in the picture of
quark and gluon configurations, which is approximated by the domain model, 
the chiral angle $\alpha$ is taken to be the 
same for all domains in the representation Eq.~(\ref{partf}).
This is a refinement of the formulation presented
in \cite{NK2001,NK2002,NK2004}. 
The measure of integration over parameters characterising domains is 
\begin{eqnarray}
\label{measure}
\int\limits_{\Sigma}d\sigma_i\dots & = & \frac{1}{48\pi}
\int_V\frac{d^4z_i}{V}
\int\limits_0^{2\pi}d\varphi_i\int_0^\pi d\vartheta_i\sin\theta_i
\nonumber \\
&\times&
\int_0^{2\pi} d\xi_i
\sum\limits_{l=0,1,2}^{3,4,5}
\delta(\xi_i-\frac{(2l+1)\pi}{6})
\nonumber\\
&\times&\int_0^\pi d\omega_i\sum\limits_{k=0,1}\delta(\omega_i-\pi k)
\dots ,
\end{eqnarray}
where $(\theta_i,\varphi_i)$ are the spherical angles of the 
chromomagnetic field, $\omega_i$ is the angle between chromoelectric and 
chromomagnetic fields and $\xi_i$ is an angle parametrising the colour 
orientation. 

This partition function describes a statistical system 
of domain-like structures of density $v^{-1}$ , where 
the volume of a domain is $v=\pi^2R^4/2$.  Each domain is
characterised by a set of internal parameters and
with internal dynamics represented by fluctuation gluon 
$Q^{(i)}$ and quark $\psi^{(i)}$ fields.
It respects the symmetries of the 
QCD Lagrangian, since the statistical 
ensemble is invariant under space-time, colour gauge 
and particularly chiral symmetries. 
Thus the model involves 
two free parameters: the mean field strength $B$ and the 
mean domain radius $R$.
These dimensionful parameters break the scale invariance
present originally in the QCD Lagrangian.
In principle, they should be
related to the trace anomaly of the energy-momentum tensor \cite{Nie77,Mink81}
and, eventually, to the fundamental scale $\Lambda_{QCD}$.
Knowledge of the full quantum effective action of QCD would be
required for establishing a relation of this kind. 
Within this framework the gluon condensate to lowest order in fluctuations
is $4B^2$ and the topological charge per domain is $q=B^2R^4/16$.

An area law is obtained for static quarks. 
The reason for this is the finite range of gluon correlations
implicit in the model which will figure in all the phenomena
we consider. 
Computation of the Wilson loop for a circular contour of
a large radius $L\gg R$ gives a string tension $\sigma = B f(\pi B R^2)$
where $f$ is given for colour $SU(2)$ and $SU(3)$ in \cite{NK2001}. 
Estimations of the values of these quantities are known from 
lattice calculation or phenomenological approaches and can be used to fit  
$B$ and $R$. As described in \cite{NK2001} these parameters are fixed to be
$\sqrt{B} = 947 {\rm {MeV}}, R=(760 {\rm{MeV}})^{-1} = 0.26 \ {\rm {fm}}$ 
with the average absolute value of topological charge per domain
turning out to be $q= 0.15$ and the density of domains 
$v^{-1}=42{\rm fm}^{-4}$. The topological susceptibility then
turns out to be $\chi = (197 {\rm MeV})^4$, comparable to the 
Witten-Veneziano value \cite{largeNc}. 
The eigenvalue problem 
\begin{eqnarray}
\!\not\!D\psi(x)&=&\lambda \psi(x),
%\nonumber\\
%i\!\not\!\eta(x) e^{i\alpha\gamma_5}\psi(x)&=&\psi(x), \ x^2=R^2
\label{Diracprob}
\end{eqnarray}
with boundary condition Eq.(\ref{quarkbc}) was studied in \cite{NK2002}.
With the domain background field the solution
to this problem reveals an asymmetric spectrum, exhibiting
the broken chiral symmetry through the bag-like boundary condition,
and thus none of the eigenmodes is chiral. However
at the centre of domains all
modes are chiral and the sign of their chirality depends on
whether the underlying gauge field is self-dual or
anti-self-dual. In \cite{NK2002} we computed the distribution of
values of the local chirality parameter of \cite{Horvath} 
in a chirally symmetric ensemble revealing qualitatively similar
behaviour to the double-peak structure seen on the lattice \cite{Xlat},
which is taken to be indicative of spontaneously broken chiral
symmetry.

The $\alpha$-dependent part of the free energy density was computed in 
\cite{NK2004} using zeta function regularisation, with an imaginary part  
arising
\begin{eqnarray}
\Im{F}= \pm  \frac{q}{v} \, {\rm arctan}(\tan(\alpha)) 
\label{anomaly}
\end{eqnarray}
where $q$ is the absolute value of topological charge in a domain, and 
overall sign $(-)+$ corresponds to an (anti-)self-dual domain.
The charge is not integer here in general but the anomalous term is 
nonetheless $\pi n$ periodic in $\alpha$.
This is the Abelian anomaly as observed
within the context of bag-like boundary conditions by
\cite{WD94} and is consistent with \cite{Fuj80} albeit
not generated from purely chiral zero modes but the chiral properties of 
non-zero modes.

The part of the free energy 
density ${\cal F}$  relevant for the present consideration of
an ensemble of $N\to\infty$ domains
with both self-dual and anti-self-dual configurations takes the form 
\begin{eqnarray*}
e^{-vN{\cal F}}&=&{\cal N}\int_{-\infty}^{\infty}d\alpha
\left[\cos v\Im F(\alpha) \right]^N
\nonumber \\
&=&{\cal N}\sum_{\alpha\in\alpha_{\rm min}}\exp\left(N
\ln[\cos(v\Im F(\alpha))]\right) + O(1/N) .
\nonumber
\end{eqnarray*} 
Here summation goes over the infinite set $\alpha_{\rm min}$ of  
degenerate minima of the free energy density, which are achieved at 
\begin{eqnarray*}
v\Im F(\alpha_{\rm min})=0({\rm mod} \ 2\pi).
\end{eqnarray*}
In the thermodynamic limit $N\to\infty$ the solutions 
to the above equation for $\alpha_{\rm min}$ 
correspond to a degenerate set of vacua connected by  
discrete chiral 
%$U_{\rm A}(1)$ 
transformations,
which will be given in the next section. 

Thus for massless quarks, 
a discrete subgroup of $U_A(1)$,  rather than the continuous $U_A(1)$ 
itself,
represents a symmetry of the vacua. 
The anomaly defines those
chiral angles which minimise the free energy
so that the full $U_A(1)$ group is no longer reflected in
the vacuum degeneracy.
It should be stressed here that this residual discrete degeneracy 
ensures a zero value for the quark condensate in the absence of mass term
or some other external chirality violating sources.

In the presence of an infinitesimally small quark mass 
the $\alpha$ dependent part of the free energy of 
a self-dual domain is modified by the term linear in mass 
and  takes the form (for details  see~\cite{NK2004})
\begin{eqnarray}
F&=&  i \frac{q}{v}{\rm arctan}[\tan(\alpha)] 
-m \aleph e^{ i \alpha } + O(m^2),
\nonumber\\
\aleph
&=&\frac{1}{\pi^2 R^3}
\sum_{k=1,z=z_1,z_1,z_2}^\infty \frac{k}{k+1}
\left[
M(1,k+2,z)
\right. \nonumber \\
&&\left. -\frac{z}{k+2}M(1,k+3,-z)-1
\right],
\end{eqnarray}
where the quantity $\aleph>0$ comes from  
the spectral asymmetry term $\eta(1)$,
\begin{eqnarray*}
\eta(s) =\sum_{\lambda} {\rm{sgn}}(\lambda)/|\lambda|^s,
\end{eqnarray*}
appearing due to the asymmetry of the spectrum~\cite{DGS98}. 
The free energy of an anti-self-dual domain
is obtained via complex conjugation. 

%%%%%%%%%%%%%%%%%%%%%%%%%%%%%%%
%%%%%%%%%%%%%%%%%%%%%%%%%%%%%%%
%SCALAR CONDENSATE FOR DOMAIN MODEL    N_F=1      q=.15
%%%%%%%%%%%%%%%%%%%%%%%%%%%%%%%
%%%%%%%%%%%%%%%%%%%%%%%%%%%%%%%
\begin{figure}[htb]
%\vspace{30mm}
\includegraphics{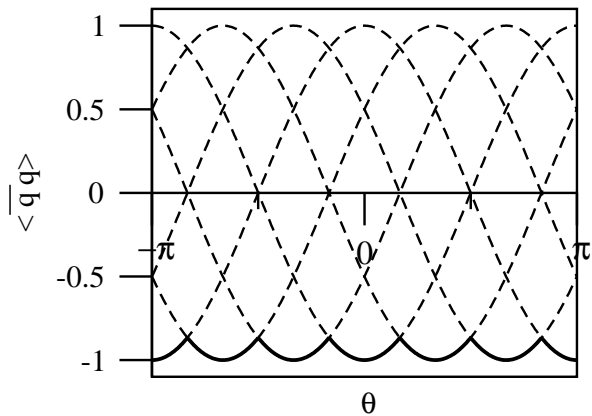}
\caption{The scalar quark condensate as a function of
the parameter $\theta$ for $N_f=1$ and $q=.15$
in units of $\aleph$.
The dashed lines correspond to discrete minima of the free energy density
which are degenerate for $m\equiv0$. The solid line denotes the minimum
which becomes global for a given $\theta$ in the presence of
infinitesimally small mass term. }
\label{fig:SNf1q}
\end{figure}
%%%%%

This discrete chiral symmetry of the massless case is spontaneously broken and 
switching on the quark mass selects one of the vacua.  
For our conventions of boundary
condition and mass term the selected minimum is at $\alpha=0$. 
It is important that values for the angles $\alpha_{\rm min}$
are not modified by the leading order term, linear in $m$.   
The quark condensate is extracted from the free energy in the standard way
\begin{eqnarray}
\langle \bar\psi(x)\psi(x)\rangle &=& 
-\lim_{m\to0}\lim_{N\to\infty} \frac{1}{vN} \frac{d}{dm}e^{-vN{\cal F}(m)}
\nonumber \\
 &=& - \aleph
\label{scalcondensate}
\end{eqnarray}
and takes the value
$\langle \bar \psi(x) \psi(x)\rangle  = -(237.8 \ {\rm MeV})^3$ 
for the values of field strength $B$ and domain radius $R$ 
fixed earlier by consideration of the pure gluonic 
characteristics of the vacuum -- string tension, topological 
succeptibility and gluon condensate.

For $N_f>1$  quark flavours the fermion boundary
condition in Eq.(\ref{quarkbc}),
explicitly breaks all chiral symmetries, flavour singlet
and non-singlet (see also \cite{WD94}). 
Thus integrating
over all $\alpha$ does not suffice to provide for the full chiral
invariance of the  ensemble of quark configurations  
contributing to the partition function. 
Rather, the boundary
condition must be generalised to include flavour non-singlet
angles, $\alpha \rightarrow \alpha +\beta^a T^a/2$,
with $T^a$ the $N_f^2-1$ generators of $SU(N_f)$. 
Then integration over $N_f^2$ angles 
$\alpha$ and $\beta^a$ must be performed
for a fully chiral symmetric ensemble. The spectrum of the Dirac
operator can be found now  quite analogously to the one-flavour case, 
except that the
boundary condition mixes flavour components and an additional
projection into flavour sectors is required in order to 
solve Eq.(\ref{Diracprob}).   
The calculation from this point
will be repeated for non-zero $\theta$ in the next
section, but it suffices to summarise here the
result for $\theta=0$ emerging in \cite{NK2004}: 
the Abelian nature of the anomaly meant that the non-singlet
chiral angles implicit in this procedure  drop out of
the ensemble free energy. Thus there are only $N_f^2-1$ continuous
directions in the vacuum. The expectation that
there should be only $N_f^2-1$ pseudo-Goldstone bosons  
 has been verified in \cite{NK2004}
for $N_f=3$ by an estimation of the meson spectrum. 

%%%%%%%%%%%%%%%%%%%%%%%%%%%%%%%
%%%%%%%%%%%%%%%%%%%%%%%%%%%%%%%
%SCALAR CONDENSATE FOR DOMAIN MODEL    N_F=3      q=.15
%%%%%%%%%%%%%%%%%%%%%%%%%%%%%%%
%%%%%%%%%%%%%%%%%%%%%%%%%%%%%%%
\begin{figure}[htb]
%\vspace{8mm}
\includegraphics{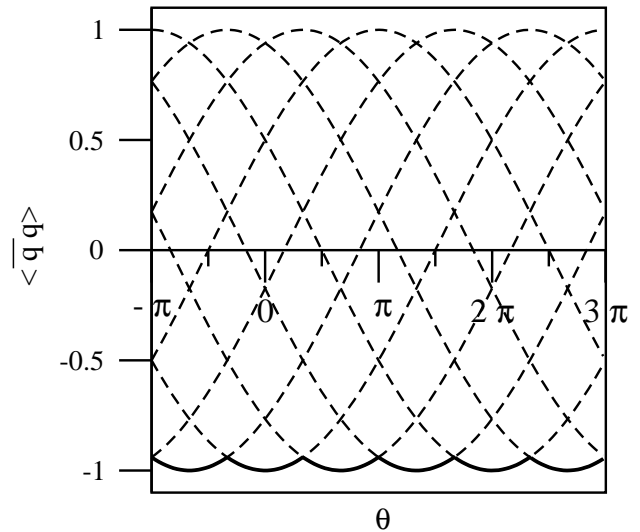}
\caption{The scalar quark condensate as a function of $\theta$
for $N_f=3$ and $q=.15$ in units of $\aleph$.
The meaning of dashed and solid lines is the same as in
Fig.~\ref{fig:SNf1q}. Qualitatively the picture is the same as
for  $N_f=1$: the same periodicity and critical values of $\theta$, which is
achieved by increased number of different discrete minima. }
\label{fig:SNf3q}
\end{figure}
%%%%%

\section{Free energy in the presence of non-zero $\theta$}
Now we include the CP-violating parameter in the model
by the additional term Eq.(\ref{thterm}) in the action,
which contributes a pure phase to the weight factor in
Euclidean space. 
Integrating over $N_f$ fermions with infinitesimally 
small masses $m_1=\dots=m_N\equiv m$ in a domain 
ensemble gives for the free energy density 
{\it per} domain
\begin{eqnarray}
\label{freeen}
F = - v^{-1}\ln\cos q[W_{N_f} -\theta] - m\aleph \sum_{i=1}^{N_f} \cos{\Phi_i},
\end{eqnarray}
 where 
\begin{eqnarray}
\label{arctantan}
W_{N_f} &=&  \sum_{i=1}^{N_f}{\rm arctan }({\rm tan}\Phi_i)
\\ 
\Phi_i&=&\alpha+B_i \nonumber \\ 
B_1 &=& 0 \ \ {\rm for} \ \ N_f=1,
\label{PhiNf1} \\
B_1&=&\frac{|\vec\beta|}{2}, \ \ B_2=-\frac{|\vec\beta|}{2}, 
\ \ {\rm for} \ \ N_f=2,
\label{PhiNf2}
\end{eqnarray}
and 
\begin{eqnarray}
B_1&=&b^3+b^8/\sqrt{3}, \ \ B_2= -b^3+b^8/\sqrt{3}, 
\nonumber \\
B_3&=&-2 b^8/\sqrt{3}, \ \ {\rm for} \ \ N_f=3.
\label{PhiNf3}
\end{eqnarray}
 Here $\alpha$ is the $U(1)$ chiral angle,  
$\vec \beta$ are the non-singlet chiral angles for $N_f$=2, 
and $b_3$ and $b_8$ are certain functions of the eight non-singlet 
chiral angles $\beta^a$ for $N_f$=3.
For $N_f=1$, Eq.(\ref{PhiNf1}) is just the result discussed in the previous
section. For $N_f=2$ the quantities $B_1$ and $B_2$
in Eq.(\ref{PhiNf2}) arise from the projection of the quark boundary condition
into $SU(2)$ flavour sectors.
For $N_f>2$ the analogous functions are $B_1, \dots, B_{N_f}$.
 For any number of flavours $N_f$
the functions $B_i$ have the property that  
\begin{equation}
\sum_{i=1}^{N_f} B_i = 0 
\label{traceless}
\end{equation}
which manifests the tracelessness of
the flavour generators in any basis. Thus  
\begin{equation}
\sum_{i=1}^{N_f} \Phi_i = N_f \alpha . 
\end{equation}
The ``$\rm{arctan}\tan$'' structure  in Eq.~(\ref{arctantan})
manifests the periodicity of the free energy for arbitrary
topological charge $q$. 
We could write 
\begin{eqnarray}
W_{N_f}=\sum_{i=1}^N {\rm{arctan}}[\tan \Phi_i] = N_f\alpha \, ({\rm mod} \ \pi)
\label{arctanresolve}
\end{eqnarray}
which is independent of the $\beta^a$. This is
the Abelian property of the anomaly 
leading to the expectation of $N_f^2-1$ Goldstone bosons.

%%%%%%%%%%%%%%%%%%%%%%%%%%%%%%%
%%%%%%%%%%%%%%%%%%%%%%%%%%%%%%%
%PSEUDO-SCALAR CONDENSATE FOR DOMAIN MODEL    N_F=1      q=.15
%%%%%%%%%%%%%%%%%%%%%%%%%%%%%%%
%%%%%%%%%%%%%%%%%%%%%%%%%%%%%%%
\begin{figure}[htb]
%\vspace{30mm}
\includegraphics{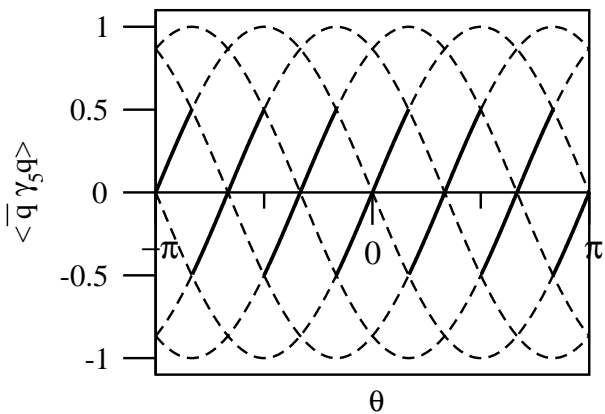}
\caption{The pseudo-scalar quark condensate as a function of
$\theta$ for $N_f=1$ and $q=.15$ in units of $\aleph$.
The meaning of dashed and solid lines is the same as in Fig.~\ref{fig:SNf1q}. }
\label{fig:PNf1q}
\end{figure}

The central effect is the chiral abelian anomaly contribution denoted
as $qW_{N_f}$ under the cosine in Eq.(\ref{freeen}).
It should be noted that 
\begin{eqnarray}
qW_{N_f}=N_f q \alpha +O(\alpha^2)\ \ {\rm for} \ \ \alpha\to 0,
\end{eqnarray}
which coincides with the standard form of axial anomaly in the 
Fujikawa derivation (Note that $\alpha$ here is twice the angle of chiral 
transformation used in the Fujikawa's calculation of anomaly \cite{Fuj80}).

The mass term is introduced as a small perturbation violating
explicitly the chiral symmetry of the system.
In the absence of the mass term, the minima of the free energy
are determined by the solutions 
\begin{eqnarray}
\label{alkl1}
\alpha_{kl}(\theta)=
\frac{\theta}{N_f}+\frac{2\pi l}{qN_f}+\frac{\pi k}{N_f} \ \ (k\in Z, l\in Z),
\end{eqnarray}
to the equation
\begin{equation}
\label{eqmin}
\cos(q W_{N_f}-q\theta)
=1.
\end{equation}
Evidently there are multiple solutions arising from the various
periodic functions appearing in Eq.(\ref{freeen}).
The index $k$ in the solutions reflects periodicity of $\tan$ 
while $l$ corresponds to the periodicity of $\cos$ in Eq.~(\ref{eqmin}).
Note that these solutions do not depend on the flavour 
non-singlet chiral angles, as discussed above,  
and thus the free energy
displays continuous degeneracy with repect to these angles.

%%%%%%%%%%%%%%%%%%%%%%%%%%%%%%%
%%%%%%%%%%%%%%%%%%%%%%%%%%%%%%%
%PSEUDO-SCALAR CONDENSATE FOR DOMAIN MODEL    N_F=3      q=.15
%%%%%%%%%%%%%%%%%%%%%%%%%%%%%%%
%%%%%%%%%%%%%%%%%%%%%%%%%%%%%%%
\begin{figure}[htb]
%\vspace{30mm}
\includegraphics{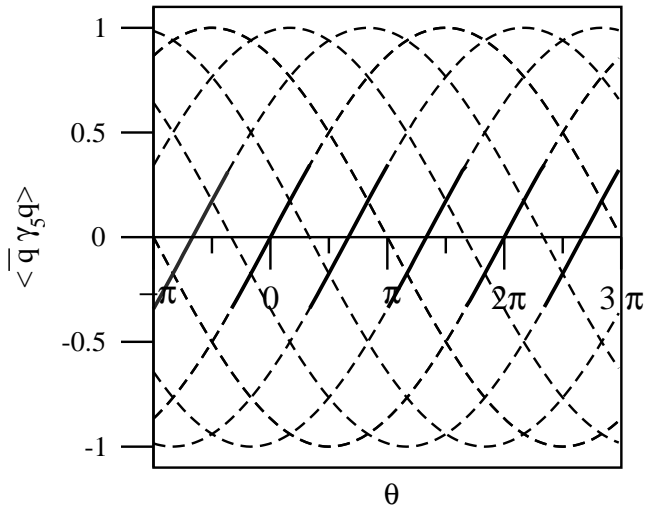}
\caption{The pseudo-scalar quark condensate as a function of
$\theta$ for $N_f=3$ and $q=.15$ in units of $\aleph$.
The meaning of dashed and solid lines is the same as in Fig.~\ref{fig:SNf3q}. }
\label{fig:PNf3q}
\end{figure}
%%%%%

The solutions Eq.~(\ref{alkl1}) give an infinite discrete set of 
degenerate minima of the free energy density.  
For given $N_f$ and $q$ a finite subset of these minima can be extracted
such that all other minima are $2\pi$-periodic copies of one of the 
already identified vacua.
For instance, for $N_f=1$ and $q=.15$ the six different vacua correspond to all 
combinations of $k=0,1$ and $l=0,1,2$, while for   
$N_f=2$ set of ``different'' minima is given by combinations of  
$k=0,1,2,3$ and $l=0,1,2$.   
It is clear that the system of minima is $2\pi$-periodic
with respect to $\theta$:
\begin{eqnarray}
\alpha_{kl}(\theta+2\pi m)\to \alpha_{k^\prime l}(\theta), \ \ k^\prime=k+2m,  
\end{eqnarray}
and the shift in $\theta$  can be
undone by re-enumerating the infinite set of minima.
This periodicity does not depend on the value of $q$ at all since it is 
ensured by re-enumerating of index $k$ without use of $l$. 
All minima are degenerate (the free energy is equal to zero)
for zero quark mass.

Switching on the infinitesimal mass term in Eq.(\ref{freeen})
lifts the degeneracy due to different values of the symmetry breaking term 
\begin{eqnarray}
\label{massterm}
V_m =-m \aleph \sum_i^{N_f} \cos(\Phi_i^{kl}(\theta))
\end{eqnarray}
in the free energy   
for different $\Phi_i^{kl}=\alpha_{kl}(\theta)+B_i(\beta^a)$.   
This introduces non-singlet angle dependence into
the free energy,  which must now
also be minimised with respect to the $\beta^a$. To this end
we expand $V_m$ in small fluctuations in the $B_i$ and
look for minima, bearing in mind the constraint Eq.(\ref{traceless}). 
The condition that $V_m$ have stationary
points leads to the condition
\begin{equation}
\sin \Phi_i \ \rm{independent \ of \ i}
\label{statpt}
\end{equation}
while the condition that this leads to a minimum forces
\begin{equation}
\cos\Phi_i  > 0  .
\label{massminimum}
\end{equation}
Eq.(\ref{statpt}) leads to $\Phi_i=\alpha+B_i$ being independent
of $i$ which means $B_i = 0 \ ({\rm{mod}} \ 2\pi)$.
On the other hand, Eq.(\ref{massminimum}) is fulfilled by
restricting $-\pi/2 \leq \Phi_i \ ({\rm{mod}} \ 2\pi) \leq \pi/2$.
However, we observe that for these values of $\Phi_i$
\begin{eqnarray*}
{\rm{arctan}}( \tan \Phi_i )= \Phi_i \ ({\rm{mod}} \ 2\pi).  
\end{eqnarray*}
Thus to guarantee a minimum with respect to
non-singlet angles, we must have
\begin{eqnarray*}
W_{N_f} &=& \sum_{i=1}^{N_f} {\rm{arctan}} (\tan \Phi_i) 
\nonumber \\
        &=& N_f (\alpha \ {\rm{mod}} \ 2\pi). 
\end{eqnarray*} 
Comparing to Eq.(\ref{arctanresolve}), we
see that of the vacua of the massless free energy, $\alpha_{kl}(\theta)$,
only those with $k$ an {\it even} integer correspond to true
minima with the mass term switched on. Note that these
arguments only apply for $N_f>1$.   For the single flavour
case the available values are $k=0,1$. A similar consideration
shows that for rational $q=n_1/n_2$ the minima selected
by the mass term correspond to $l$ any integer if $n_1$ is odd
but $l$ even if $n_1$ is even.   

The result is that the $\theta$ dependence of the quark
condensate is given by
\begin{eqnarray*}
\langle {\bar{\psi}}(x) \psi(x) \rangle _{\alpha_{kl}(\theta)}=- N_f\aleph \cos(\alpha_{kl}(\theta))
\end{eqnarray*}
for $k$ even and $l$ any integer except if $q=n_1/n_2$ with $n_1$ even
for which then $l$ is also even.
This is plotted in Figs.~\ref{fig:SNf1q} and  \ref{fig:SNf3q}
for $N_f=1$ and $N_f=3$ respectively,
where the cosine in Eq.(\ref{massterm}) is plotted as a function of $\theta$
for the minima which are not $2\pi$-equivalent.
Critical values of $\theta$ correspond to the points where two minima
become degenerate.

The two degenerate vacua at such critical points are distinguished
by their CP properties, which can be seen from the behaviour of
pseudoscalar condensate as a function of $\theta$
\begin{eqnarray}
\langle {\bar \psi}(x) \gamma_5 \psi(x) \rangle_{\alpha_{kl}(\theta)}=-N_f\aleph\sin(\alpha_{kl}(\theta)),
\nonumber
\end{eqnarray}
as derived for the domain model in Appendix~A.
As expected, the scalar (see Eq.~(\ref{massterm}))
and pseudo-scalar condensates
depend on $\theta$ through cosine and
sine of $\alpha_{kl}(\theta)$ respectively.

The plots for one and three flavours respectively are given
in Figs.~\ref{fig:PNf1q} and  \ref{fig:PNf3q} again with $q=0.15$.
The pseudoscalar condensate is discontinuous at
the critical values of $\theta$ and takes values opposite in sign for the
two degenerate minima:
parity is thus spontaneously broken.
                                                                                
In other words, we can see that for most values of theta the mass term selects aunique minimum of the free energy.
However, there are critical values of $\theta$ where two different minima
are degenerate, thus displaying a two-fold degeneracy of the vacuum in the
presence of a mass.

There are two conditions
for critical $\theta$. The first one is obviously
\begin{eqnarray}
\alpha_{00}(\theta_{\rm{crit}})=-\alpha_{kl}(\theta_{\rm{crit}}) \  ({\rm mod} \ 2\pi)
\nonumber
\end{eqnarray}
where $k$ and $ l$ should not be equal to zero simultaneously and without
loss of generality we have taken $\alpha_{00}$ on the RHS.
Thus
\begin{eqnarray}
\theta^{kl}_{\rm{crit}}=\pi k + \frac{l\pi}{q} \ ({\rm mod} \ 2\pi) .
\nonumber
\end{eqnarray}
For a given $q$ this defines set of values of $\theta^{kl}_{\rm crit}$ where
several vacua are degenerate.
This set is independent of number of flavours $N_f$.
Furthermore, we are interested only in those $\theta^{kl}_{\rm crit}$
which minimise the term linear in mass, which is
the second condition for $\theta_{\rm crit}$.
It is easy to check that independently of $q$ the value
$\theta_{\rm crit}=\pi$ satisfies both conditions.
Other critical values depend on topological charge of the domain $q$.
In general if $q=n_1/n_2$ then there are $n_1$ critical values of theta.
For the value $q=0.15$, as was fit in the domain model, we find
\begin{eqnarray*}
&&\theta_{\rm crit}=\left\{-\pi/3,\pi/3,\pi\right\} \ ({\rm mod} \ 2\pi),
\end{eqnarray*}
where the Dashen phenomenon occurs.
We conclude that for any $N_f$ and rational $q$ there is a finite number of
critical points in the interval
$[0,2\pi]$ including $\theta=\pi$.

\section{Summary and discussion}

The central result of this paper is that
$2\pi$-periodicity of amplitudes in $\theta$ and vacuum doubling with
spontaneous CP violation at certain critical values of $\theta$,
in particular $\theta=\pi$,
are obtained for any number of light flavours $N_f$ and arbitrary
rational topological charge $q$. This is achieved
in a model whose input is a particular class of
nonperturturbative gluon configurations. From this model
have been derived both confinement, the correct
pattern of chiral symmetry breaking,
certain static characteristics of the
vacuum (string constant, condensates, topological susceptibility)
as well as properties of the meson spectrum
(correct splitting between masses of iso-vector and iso-singlet states).
The most important qualitative feature of this class of
vacuum fields is that it can be seen as an ensemble of densely packed lumps
of (anti-)self-dual gluon fields, characterised by a finite correlation length.
In this paper and in previous works it has been shown on a semi-quantitative
level that such vacuum fields
can reproduce all of the qualitatively important features of the QCD vacuum
associated with confinement and chiral symmetry realisation.

%%%%%%%%%%%%%%%%%%%%%%%%%%%%%%%
%%%%%%%%%%%%%%%%%%%%%%%%%%%%%%%
%Crewther's SCALAR CONDENSATE    N_F=3      q=1/N_F
%%%%%%%%%%%%%%%%%%%%%%%%%%%%%%%
%%%%%%%%%%%%%%%%%%%%%%%%%%%%%%%
\begin{figure}[htb]
%\vspace{8mm}
\includegraphics{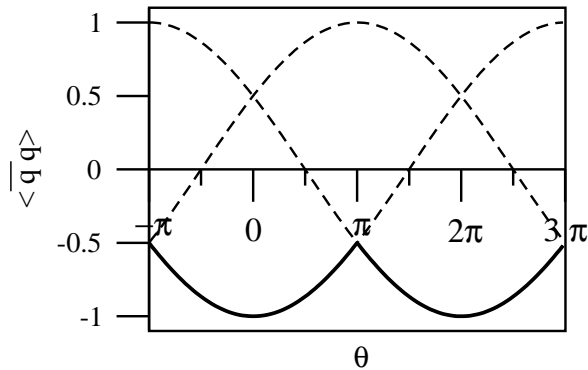}
\caption{The scalar quark condensate as a function of $\theta$ for
$N_f=3$ and $q=1/N_f$ in units of $\aleph$,  which is equivalent
to corresponding plot from \cite{CrewNATO}.
The meaning of dashed and solid lines is the same as in Fig.~\ref{fig:SNf1q}.}
\label{fig:SNf3}
\end{figure}
%%%%%

We discuss now our results in light of the two main approaches, that
of anomalous Ward identities and of 
the effective chiral Lagrangian at large $N_c$, which
originally predicted the Dashen phenomenon in the
presence of a theta term. For convenience, we
have summarised the salient features of these approaches 
in Appendix~B.

The first point of comparison is the minimisation with respect
to non-singlet angles. In both \cite{CrewNATO} and \cite{largeNc},
this is referred to as ``Dashen's theorem''. In the
domain model context, precisely the same conditions 
Eq.(\ref{statpt},\ref{massminimum}) 
have arisen by considering the free energy as a function
of the dynamical variables $\beta^a$ associated with
the flavour non-singlet angle dependence of fermion
boundary conditions.  
In \cite{CrewNATO} and \cite{largeNc}
the corresponding angles are denoted $\phi_i$ corresponding to the phases
of an $N_f\times N_f$ matrix, either the
matrix of scalar condensates $\langle \bar{\psi}_i\psi_j\rangle$
in \cite{CrewNATO} or the chiral field of \cite{largeNc}.
In all cases, these results (for the degenerate mass case)
emerge because of the
structure of the symmetry breaking term $-m\aleph \sum_{i=1}^{N_f}\cos\Phi_i$.
The significance of this structure for the domain model is
that it emerges precisely from the 
spectral asymmetry of the Dirac operator in
the domain field background as
computed via zeta function regularisation \cite{NK2002}. 
The analogues of the $\phi_i$ in \cite{CrewNATO,largeNc} are
our functions $\Phi_i$ associated with the flavour dependence
of the fermion domain boundary conditions.

The second point of contact relates to the
existence or otherwise of an axial $U(1)$ Goldstone boson
in the spectrum. In both \cite{CrewNATO,largeNc},
the {\it input} that there be no such boson leads to
a constraint on the sum of the aforementioned angles $\phi_i$
and the theta parameter:
\begin{equation}
\sum_{i=1}^{N_f} \phi_i = \theta .
\label{noU1boson}
\end{equation}
In the domain model approach, as mentioned at the
outset, the anticipated absence of such a boson is a {\it consequence}   
of the lack of continuous axial $U(1)$ degeneracy of the ground state
of the free energy for the massless case. But this ground state is
determined by 
\begin{equation}
\sum_{i=1}^{N_f} \arctan(\tan \Phi_i) = \theta
\label{ourvacuum}
\end{equation}
from Eq.(\ref{eqmin}). Resolving the ${\rm{arctan}}\tan$
structure brings us to the same form as in Eq.(\ref{noU1boson}),
when $\phi_i$ and $\Phi_i$ are identified,
again consistent with our observations above.

%%%%%%%%%%%%%%%%%%%%%%%%%%%%%%%
%%%%%%%%%%%%%%%%%%%%%%%%%%%%%%%
%Crewther's PSEUDO-SCALAR CONDENSATE    N_F=3      q=1/N_F
%%%%%%%%%%%%%%%%%%%%%%%%%%%%%%%
%%%%%%%%%%%%%%%%%%%%%%%%%%%%%%%
\begin{figure}[htb]
%\vspace{8mm}
\includegraphics{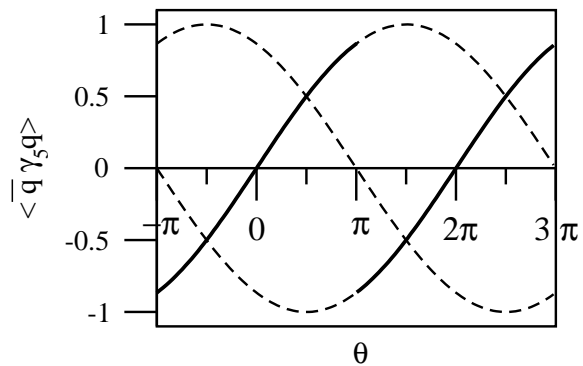}
\caption{The pseudo-scalar quark condensate as a function of $\theta$ 
for $N_f=3$ and $q=1/N_f$ in units of $\aleph$.  
The meaning of dashed and solid lines is the same as in Fig.~\ref{fig:SNf1q}.}
\label{fig:PNf3}
\end{figure}
%%%%%

Certainly the reason for these coincidences is most 
transparent in the comparison to the large $N_c$ approach,
reviewed in appendix B: the effective singlet and non-singlet 
meson fields there
are contained in $U(N_f)$ matrices but diagonalised
by $SU(N_f)\times SU(N_f)$ transformations. In
the domain model, where the conditions on quark
fields at domain boundaries also involve the $U(N_f)$ 
flavour matrices $\exp[i (\alpha + \beta^a T^a/2) \gamma_5]$
which are similarly diagonalised by special unitary matrices.  

The main difference with \cite{CrewNATO} is the specific
value of the topological charge. For a more direct
comparison with \cite{CrewNATO} we set $q=1/N_f$ in our
formula, Eq.(\ref{alkl1}). The $l-$dependent term is
then $2\pi l$ which is inconsequential under the cosine
for the condensate.  Thus our results completely agree with \cite{CrewNATO} for this case.
For the $N_f=3$ case, we plot the scalar and pseudoscalar
condensates in  
Figs.~\ref{fig:SNf3} and \ref{fig:PNf3} which are identical to corresponding plots in \cite{CrewNATO}.

The significance of our result in
comparison to \cite{CrewNATO,largeNc} is that the domain model
assumes the dominance of specific class of nonperturbative gluon 
configurations in the QCD functional integral and thereby provides 
the $2\pi$ periodicity in theta
dependence and the existence of critical values of theta parameter as output 
simultaneous with the
correct resolution of the axial $U(1)$ problem.
In \cite{CrewNATO,largeNc} the absence of a pseudoscalar
$U(1)$ boson is input in order to obtain the theta dependence,
while the responsible vacuum structure is unknown.

The present work can be generalised for the case of non-degenerate
quark masses in the presence of the theta term
for which an analysis using anomalous Ward
identities is given in \cite{CrewNATO}. Because of the
above-noted similarities with the domain model  
the condensate dependence on $\theta$ 
will be identical.
 
Finally, we mention that the model under consideration 
with rational topological charges $q$ reflects the strong 
CP problem in the usual way,   
but certainly cannot resolve it. 
As in several other approaches
(for instance~\cite{Mink}) we notice that the  
free energy in the model is minimised by $\theta=0$,
but there is no reason within the model to
demand the minimisation of the free energy 
with respect to $\theta$, which is an external parameter here.
However, allowing irrational values of $q$ can drastically change
the status of the strong CP problem in the model due to 
the appearance of infinitely many
critical values of $\theta$ in any finite interval.  
We shall analyse this intriguing possibility in a separate publication.

\section*{Acknowledgements}
ACK is supported by the Australian Research Council.
SNN was supported by the DFG, contract SM70/1-1
and partially by the grant RFBR~04-02-17370.
ACK thanks R.J. Crewther and B.-Y.Park for helpful discussions. 

\appendix
\section{Pseudo-scalar condensate}

We treat here $N_f=1$  and the integral for one domain
for simplicity; the generalisation is straightforward.
Consider
\begin{eqnarray*}
\langle  \bar\psi   e^{i\beta\gamma_5} \psi\rangle_\alpha
&=& 
\frac{1}{2}\left[
\langle  \bar\psi   e^{i\beta\gamma_5} \psi\rangle^{+}_\alpha+
\langle  \bar\psi   e^{i\beta\gamma_5} \psi\rangle^{-}_\alpha
\right]
\\
\langle  \bar\psi   e^{i\beta\gamma_5} \psi\rangle^{\pm}_\alpha
&=& \lim_{m\to 0}v^{-1}Z^{\pm}_\alpha(m,\beta)\\
Z^{\pm}_\alpha(m,\beta)&=&{\cal N}\int_{{\cal F}_{\alpha}}D\psi D\bar\psi
\exp\left\{- S \right. \nonumber \\
&& \left. + m\int_v dx \bar\psi e^{i\beta\gamma_5}\psi   \right\},
\end{eqnarray*}
where $(-)+$ corresponds to an (anti-)self-dual domain.
A chiral transformation:
\begin{eqnarray}
\label{chtr}
\psi=e^{-i\frac{\beta}{2}\gamma_5}\psi^\prime
\end{eqnarray}
for a field $\psi$ belonging to ${\cal F}_{\alpha}$, 
namely satisfying 
\begin{eqnarray*}
i\!\not\!\eta(x)e^{i\alpha\gamma_5}\psi(x)=\psi(x),
\end{eqnarray*}
leads to the transformed field $\psi^\prime$ satisfying 
 \begin{eqnarray*}
i\!\not\!\eta(x)e^{i(\alpha-\beta)\gamma_5}\psi^\prime(x)=\psi^\prime(x),
\end{eqnarray*}
so that $\psi^\prime$ belongs to ${\cal F}_{\alpha-\beta}$.
Performed in the integral, this chiral transformation results in
\begin{eqnarray}
\label{phase1}
Z^{\pm}_\alpha(m,\beta)=e^{i\Gamma(\beta)}Z^{\pm}_{\alpha-\beta}(m,0)
\end{eqnarray}
where the phase $\Gamma$ 
is fixed by $\beta$-independence of  $Z^{\pm}_\alpha(0,\beta)$
and the result for the fermionic determinant 
\begin{eqnarray*}
&&Z^{\pm}_\alpha(0,0)=\exp(\pm  i q\ {\rm arctan}[{\rm tan}(\alpha)]),
\end{eqnarray*}
which gives 
\begin{eqnarray}
\label{phase2}
\Gamma(\beta)=\pm 2  \ {\rm arctan}[{\rm tan}(\beta)].
\end{eqnarray}
Thus, collecting all together, we get 
\begin{eqnarray*}
Z^{\pm}_\alpha(m,\beta)&=&\exp\left\{\pm  i q\ {\rm arctan}[{\rm tan}(\alpha)]
\right. \nonumber \\
&&\left. +v m \exp(\pm i (\alpha-\beta))\aleph\right\}.
\end{eqnarray*}
In the presence of the $\theta$-term this gives
\begin{eqnarray*}
\langle  \bar\psi   e^{i\beta\gamma_5} \psi\rangle^{\pm}_\alpha
&=&-e^{\pm  i q\ ({\rm arctan}[{\rm tan}(\alpha)]-\theta)}
[ \cos(\alpha-\beta)\nonumber \\
&& \pm i  \sin(\alpha-\beta)]\aleph,
\end{eqnarray*}
which finally, after summing self-dual and anti-self-dual configuration, 
leads to
\begin{eqnarray*}
\langle  \bar\psi   e^{i\beta\gamma_5} \psi\rangle_\alpha=
-e^{\ln\cos\{ q\ ({\rm arctan}[{\rm tan}(\alpha)]-\theta)\}} 
\aleph \cos(\alpha-\beta).
\end{eqnarray*}
Finally using this result for the complete calculation with $N\to\infty$ 
domains and integration over angles $\alpha_j$ associated 
with $j$-th domain, we get 
\begin{eqnarray*}
\langle  \bar\psi   e^{i\beta\gamma_5} \psi\rangle=-\aleph \sum_{kl} \cos(\alpha_{kl}(\theta)-\beta),
\end{eqnarray*}
where the sum spans all  $2\pi$-inequivalent minima $\alpha_{kl}(\theta)$ of the free energy density,
explicitly given in Eq.~(\ref{alkl1}).

In particular, the pseudoscalar condensate corresponds to $\beta=\pi/2$ and thus reads 
\begin{eqnarray*}
\langle  \bar\psi   i\gamma_5 \psi\rangle=N_f \aleph \sum_{kl} \sin(\alpha_{kl}(\theta))
\end{eqnarray*}
It should be stressed here again, that for any $\theta$ and $\beta$
$$
\sum_{kl} \cos(\alpha_{kl}(\theta)-\beta)\equiv 0.
$$
We see that the $\theta$-dependence of the scalar condensate ($\beta=0$) for 
different minima of the free energy density
is given by $\cos(\alpha_{kl}(\theta))$ while the dependence of the pseudoscalar condensate ($\beta=\pi/2$) 
is described by $\sin(\alpha_{kl}(\theta))$,
as shown in Figs.~\ref{fig:PNf1q} and \ref{fig:PNf3q}.

The crucial point in this derivation is that the chiral transformation Eq.~(\ref{chtr})
changes the space of integration and simultaneously generates a phase $\Gamma$ in Eq.~(\ref{phase1}).
This phase is fixed in the form of Eq.~(\ref{phase2}) by substituting $m=0$ into Eq.~(\ref{phase1})
\begin{eqnarray*}
Z^{\pm}_\alpha(0,\beta)=e^{i\Gamma(\beta)}Z^{\pm}_{\alpha-\beta}(0,0),
\end{eqnarray*}
and taking into account the known explicit form of 
$Z^{\pm}_{\alpha-\beta}(0,0)$ and $Z^{\pm}_\alpha(0,\beta)$, 
where the latter does not depend on $\beta$ by construction.

\section{Other derivations of the Dashen Phenomenon}
\subsection{Anomalous Ward identities}
We summarise the salient aspects of \cite{CrewNATO}
and related works.
Denote by $J_{\mu 5}$ the singlet axial vector current renormalised
gauge-invariantly. Its divergence gives the
anomaly. Inserting this current into Green's functions
with a composite operator consisting of a product
of local observables $O_k(x_k)$, taking the divergence
and using the anomaly one obtains   
\begin{eqnarray}
&&\partial^{\mu}_x T\langle 0 | J_{\mu 5}(x) \prod_k O_k (x_k)|0 \rangle 
= \nonumber \\
&& 2N \partial^{\mu}_x T  \langle 0 | K_{\mu}(x)  \prod_k O_k (x_k)|0 \rangle
\nonumber \\
&&- \sum_l \chi_l \delta^{(4)}(x-x_l) T\langle 0 | \prod_k O_k (x_k)|0 \rangle
\label{divcurrent}
\end{eqnarray}
The quantity $K_{\mu}$ in the first term of the RHS
is the well-known Chern-Simons current arising from the
anomaly. The second term arises
from commuting the divergence through the T-product, which
generally gives a commutator of the operators $O_k$
with the axial charge, and then    
rewriting that commutator in terms
of chiralities $\chi_k$ corresponding to
$O_k$ which are defined via the eigenvalue-like 
relation 
\begin{equation}
[Q_5, O_k] = - \chi_k O_k .
\label{chiralitiesdef}
\end{equation}
The charge $Q_5$ does not correspond to $J_{\mu 5}$ but 
rather to the gauge-dependent, conserved axial current.
Despite the gauge dependence of $Q_5$,
chiralities of Eq.(\ref{chiralitiesdef}) are renormalisation group- and
gauge-invariant \cite{CrewNATO}. For a right handed quark field $\chi=1$.
The condition for avoiding a $U(1)$ boson is that the
LHS of Eq.(\ref{divcurrent}) vanishes at zero momentum
transfer. Taken between $\theta$ vacua, one can rewrite the
Chern-Simons current contribution (due to its connection
to topological charge density) on the RHS via a derivative with respect
to $\theta$ yielding  
\begin{equation}
0=[2N_fi {{\partial}\over{\partial\theta}}  - \sum_l \chi_l ]
T\langle 0 | \prod_k O_k (x_k)|0 \rangle
.
\end{equation}
Now one chooses the local operator product ${\bar q} q$
such that the sum of chiralities is $\sum_l \chi_l=2$.
One extracts then the relation
\begin{equation}
(i N_f  {{\partial}\over{\partial\theta}}  -1 ) 
\langle 0 | {\bar u}_L u_R |0 \rangle = 0
\label{condensatecondition}
\end{equation}
for example.
This can be trivially solved for the condensate.
The next step is to recognise the condensate here as an element
of the matrix of condensates which break the non-singlet
chiral symmetry
\begin{eqnarray*}
\langle 0 | ({\bar q_L})_i (q_R)_j |0 \rangle = C V_{ij}.
\end{eqnarray*}
This matrix can be brought by
chiral rotations into a form ${\rm{diag}}(e^{i\phi_i})$.
The real angles $\phi_i$ parametrising the matrix
now carry the $\theta$ dependence of the condensate.   
Inserting this into Eq.(\ref{condensatecondition})
gives 
\begin{eqnarray*}
\sum_{i=1}^{N_f} (N_f {{\partial \phi_i}\over{\partial\theta}}  -1 )=0
\end{eqnarray*}
whose solution is 
\begin{equation}
\sum_{i=1}^{N_f} \phi_i - \theta = 0. 
\label{nogoldstone}
\end{equation}
This equation should be understood here as a direct consequence
of the requirement that no zero mass boson
couple to the gauge-invariant axial vector current.
  
The above considerations should be repeated in the
presence of such a perturbation matrix of masses $m_i$ for each
flavour  
\begin{eqnarray}
\langle \epsilon H' \rangle &=& 
\sum_{i,j=1}^{N_f} m_i (V_{ij} + V^{\dagger}_{ij})
 \nonumber \\
&=& 2 \sum_{i=1}^{N_f} m_i \cos\phi_i 
\end{eqnarray}
According to Dashen's theorem, the true vacuum is found by minimising a 
quark mass term with respect
to small chiral rotations about this configuration,  
meaning shifting $\phi_i$ under the cosine by infinitesimal $\omega_i$.
Minima are determined by the conditions 
\begin{eqnarray}
m_i \sin\phi_i &=& {\rm{independent \ of}} \ i \equiv \lambda
\nonumber \\
\cos\phi_i &>&0
.
\label{Dashentheorem}
\end{eqnarray}
To give an explicit solution assume $N_f=2$ with degenerate quark masses. 
Then the consequence of Dashen's
theorem Eq.(\ref{Dashentheorem}) gives $\sin \phi_i=\lambda/m $.
Solutions consistent with Eq.(\ref{Dashentheorem}) 
are $\phi_i=\phi \in [-\pi/2,\pi/2]$
modulo $2\pi$, with $\phi=\arcsin\lambda/m$.
The absence of a Goldstone
boson Eq.(\ref{nogoldstone}) gives $\phi_1+\phi_2=\theta$. 
Combining these gives the result that the symmetry
breaking term and thus the condensate is $|\cos(\theta/2)|$.
Essentially the absolute value appears because the cosine
may not change sign, due to Eq.(\ref{Dashentheorem}), as
$\theta$ varies.
Out of this emerges that the periodicity of $\theta$ is
$2\pi$. For general $N_f$ with degenerate quark
masses the corresponding result is that  
the condensate is proportional to $\cos[\theta ({\rm{mod}} 2\pi)/N_f]$. 

\subsection{Large $N_c$ approach}
Now we briefly summarise how identical results are obtained in
the large $N_c$ approach of Witten and Veneziano-Di Vecchia
\cite{largeNc}.
An $N_f\times N_f$ unitary field $U$, parametrised as   
$U=U_0\exp(i\pi^a t^a)$ with $N_f^2$ meson (non-singlet and
singlet) fields $\pi^a$, is considered:
the $U(N_f)$ generators $t^a$ include the identity matrix as well as the usual
$SU(N_f)$ generators.  
An effective Lagrangian for $U$ with
chiral and axial $U(1)$ symmetry broken is
\begin{eqnarray*}
{\cal L} &=& \frac{F_{\pi}^2}{2} ( 
{\rm{Tr}} \partial_{\mu}U  \partial_{\mu} U^{\dagger}  
+ {\rm{Tr}} ({\cal M} U + {\cal M}^{\dagger} U^{\dagger})
\nonumber \\
&&  - \frac{a}{N_c} (-i \ln \det U)^2 ) 
\end{eqnarray*}
where the first term is a kinetic term, the second a mass,
and the third term is designed to only yield a term
quadratic in the singlet field and is consistent with  
large $N_c$ counting rules.
The mass matrix ${\cal M}$ is then diagonalised by an
$SU(N_f)\times SU(N_f)$ transformation (corresponding
to the action on the quark mass matrix) to the form
\begin{eqnarray*}
{\cal M} = e^{i\theta/{N_f}} M
\end{eqnarray*}
with $M={\rm{diag}}(\mu_i^2)$. The $\mu_i^2$
are combinations of the squares of the meson masses 
but are linear in the corresponding quark masses $m_i$,
$\mu_i^2 \propto m_i$,
generalising the Gell-Mann--Oakes--Renner relationship
and containing the quark condensates - see \cite{Pich91} for explicit formulae.
The diagonalisation leads to a corresponding phase
transformation on the $U-$field: $U\rightarrow e^{-i\theta/{N_f}} U$
such that the $\ln\det$ term undergoes a shift $\theta$.
The aim is now to minimise the effective potential to find
the vacuum configurations. This is aided by considering a 
diagonal $U$ parametrised as ${\rm{diag}}(e^{i\phi_i})$
where the $\phi_i$ are complicated functions of the
meson fields $\pi^a$. 
This leads to the potential
\begin{eqnarray*}
V(\phi_i) = F_{\pi}^2 \left[-\sum_i \mu_i^2 \cos\phi_i + 
\frac{a}{2N_c} \left(\sum_i \phi_i - \theta\right)^2  \right] .
\end{eqnarray*}
The terms with $\mu_i^2$ evidently arise from the separate
phases in $U$ and thus after expanding the $\phi_i$ in powers of 
the meson fields $\pi^a$ will give mass terms for
the $N_f^2-1$ non-singlet mesons.  
The last term incorporates only information about the
overall phase of $U$; the corresponding expansion
to second order in $\pi^a$ will reflect the mass of the singlet $U(1)$ state. 
That this is taken to be large compared to the other $N_f^2-1$
is now input at this point.  
One now minimises with respect to the angles $\phi_i$
(``Dashen's theorem'') giving 
\begin{eqnarray*}
\mu_i^2 \sin\phi_i = \frac{a}{N_c} \left(\sum_{j=1}^{N_f} \phi_j 
- \theta\right). 
\end{eqnarray*}
Evidently then one has again   
\begin{eqnarray*}
\mu_i^2 \sin\phi_i = {\rm{independent \ of}} \ i
\end{eqnarray*}
identical to Eq.(\ref{Dashentheorem}). 
On the other hand, taking the first $N_f^2-1$ mesons to be light,
the only way to minimise the potential for a heavy $U(1)$ boson
is to constrain the term with factor $a/N_c$ 
to be exactly zero, namely  
\begin{eqnarray*}
\sum_i \phi_i - \theta = 0
\end{eqnarray*}
which is identical to the condition from
anomalous Ward identities - no surprise since the effective
Lagrangian is engineered to satisfy these identities.
Implementing these conditions leads to the identical
dependence on $\theta$ discussed in the previous section, which
now appears in the potential at its minima 
$$
V(\phi_i)_{\rm{min}} =-F_{\pi}^2 \sum_i \mu_i^2 \cos\phi_i|_{\rm{min}}
.
$$
Since the quark condensates reside in the $\mu_i^2$  
an identical dependence on $\theta$ 
as that found using anomalous Ward identities emerges. 
The dependence involves $\cos(\theta/N_f)$
now because of the way the $U-$field transforms under
chiral transformations in terms of the $\theta$ parameter: 
$\ln\det U \rightarrow \ln \det U - i \theta$ and there
is no relative factor of two between the sum of the phases of $U$
and $\theta$.

\end{document}